\PassOptionsToPackage{usenames}{color}
\documentclass[a4paper]{easychair}
\usepackage[T1]{fontenc}
\usepackage{lmodern}
\usepackage{amsmath,amssymb,amsthm,mathtools}
\usepackage{setspace}
\usepackage{enumitem}
\usepackage[pdftex]{virginialake}
\xyoption{pdf}
\newcommand{\IHproof}[1]{%
  \vltr{\mathrm{IH}}{#1}
    {\vlhy{}}{\vlhy{\hspace{1.5em}}}{\vlhy{}}%
}
\providecommand{\IHcase}[2]{%
  \vltr{\mathrm{IH}_{#1}}{#2}
    {\vlhy{}}{\vlhy{\hspace{1.5em}}}{\vlhy{}}%
}
\usepackage{microtype}
\usepackage{graphicx,adjustbox,needspace,etoolbox}
\usepackage{array}
\usepackage{bookmark}
\hypersetup{
  pdftitle={The Failure of the Uniform Interpolation Property of Intuitionistic Epistemic Logic IEL},
  pdfauthor={Xin Li}
}
\allowdisplaybreaks[2]
\setlist[enumerate]{nosep}
\setlist[itemize]{nosep}
\providecommand{\tightlist}{\setlength{\itemsep}{0.15em}\setlength{\parskip}{0pt}}

\numberwithin{equation}{section}
\theoremstyle{plain}
\newtheorem{theorem}{Theorem}[section]
\newtheorem{lemma}[theorem]{Lemma}
\newtheorem{corollary}[theorem]{Corollary}
\newtheorem{definition}[theorem]{Definition}
\BeforeBeginEnvironment{theorem}{\Needspace{3\baselineskip}}
\BeforeBeginEnvironment{lemma}{\Needspace{7\baselineskip}}
\BeforeBeginEnvironment{corollary}{\Needspace{7\baselineskip}}
\BeforeBeginEnvironment{definition}{\Needspace{6\baselineskip}}

\title{The Failure of the Uniform Interpolation Property of Intuitionistic Epistemic Logic $\mathrm{IEL}$}
\author{
  Xin Li  
}
\institute{The Tsinghua-UvA JRC for Logic, Department of Philosophy, Tsinghua University, China}
\date{}
\authorrunning{~}
\titlerunning{~}

\begin{document}

\maketitle

\begin{abstract}
We prove that  neither intuitionistic epistemic logic $\mathrm{IEL}$ nor its weaker variant $\mathrm{IEL}^{-}$, obtained by omitting the weak factivity axiom $\Box\varphi\to\neg\neg\varphi$, has the uniform interpolation property. More precisely, the formula $((q\to p)\land(\Box p\to p))\land(p\to r)$ has no uniform post-interpolant, and the formula $(((q\to p)\land(\Box p\to p))\land(p\to r))\to r$ has no uniform pre-interpolant, both with respect to the same propositional variable $p$. The counterexamples work in every logic between $\mathrm{IEL}^{-}$ and $\mathrm{IEL}$. The proof is syntactic and uses the cut-free sequent calculus of Krupski and Yatmanov.
\end{abstract}

\section{Introduction}

Interpolation property concerns whether a provable implication can be factored through a formula using only the vocabulary shared by its premise and conclusion. Its classical starting point is Craig's interpolation theorem~\cite{Craig1957}. In the propositional setting, a logic $L$ has the Craig interpolation property if, whenever $\varphi\to\psi$ is derivable in $L$, there is a formula $\chi$ using only variables common to $\varphi$ and $\psi$ such that both $\varphi\to\chi$ and $\chi\to\psi$ are derivable in $L$. This formula $\chi$ is called an interpolant of $\varphi$ and $\psi$.

Uniform interpolation strengthens Craig interpolation by requiring the interpolant to depend only on the premise, or, dually, only on the conclusion of the implication once the propositional variables allowed in the interpolant are fixed. Given a formula $\varphi$ and a variable $p$, a uniform post-interpolant is a strongest $p$-free consequence of $\varphi$, whereas a uniform pre-interpolant is a weakest $p$-free sufficient premise for $\varphi$. In both cases, the interpolant must use only variables of $\varphi$ other than $p$. The requirement is uniform because the same formula must account for every $p$-free consequence or sufficient premise, even one involving additional variables. It must therefore be chosen independently of the other side of the implication. Uniform interpolation asks, in this sense, whether eliminating a variable leaves information that can still be expressed by a single formula of the original language.

Pitts~\cite{Pitts1992} proved uniform interpolation for intuitionistic propositional logic via a constructive proof-theoretic method, thereby giving an interpretation of propositional quantification within the propositional language. Subsequent work has investigated how far this kind of method extends to intuitionistic modal logics, that is, extensions of intuitionistic propositional logic with one or more modal operators. Iemhoff~\cite{Iemhoff2019} established sufficient structural conditions on sequent calculi for uniform interpolation, obtaining, in particular, the uniform interpolation property for the box-only intuitionistic versions of $K$ and $KD$. F\'{e}r\'{e}e, van der Giessen, van Gool, and Shillito~\cite{FereeEtAl2024} gave a mechanised proof for intuitionistic strong L\"ob logic $\mathrm{iSL}$. More recently, van der Giessen and Shillito~\cite{GiessenShillito2026} proved uniform interpolation for constructive $K$ and Wijesekera's $K$ with independent box and diamond operators. These results provide both positive examples and methods for constructing uniform interpolants from suitable terminating calculi.

The present paper concerns the intuitionistic epistemic logics $\mathrm{IEL}^{-}$ and $\mathrm{IEL}$, introduced by Artemov and Protopopescu~\cite{ArtemovProtopopescu2016}, to formalise intuitionistic belief and knowledge, respectively. The intended reading of $\Box\varphi$ is that $\varphi$ has been verified, where verification need not provide an explicit proof of $\varphi$. Both extend intuitionistic propositional logic with the axiom $(K)$, $\Box(\varphi \to\psi) \to (\Box \varphi \to \Box \psi)$ and co-reflection $\varphi\to\Box\varphi$, while $\mathrm{IEL}$ additionally validates weak factivity, $\Box\varphi\to\neg\neg\varphi$. Krupski and Yatmanov~\cite{KrupskiYatmanov2016} developed a cut-free sequent calculus for $\mathrm{IEL}$ and established its PSPACE-completeness. Su and Sano~\cite{SuSano2019} established Craig interpolation in their study of first-order intuitionistic epistemic logic. These results motivate the question whether $\mathrm{IEL}^{-}$ and $\mathrm{IEL}$ have uniform interpolation. We answer the question negatively by giving counterexamples.

Uniform interpolation is known to fail for several modal logics. Ghilardi and Zawadowski~\cite{GhilardiZawadowski1995} showed that it fails for classical modal logic $S4$. Van der Giessen~\cite[Section~3.4]{Giessen2022} established failure for the box-only intuitionistic logics $\mathrm{iK4}$ and $\mathrm{iS4}$. Her argument transfers classical negative results to the intuitionistic setting by means of negative translations.

Our proof is motivated by the connection between uniform interpolation and fixed-point iteration; see D'Agostino~\cite{DAgostino2007} for a broader discussion of this connection. The guiding idea is to use a fixed-point assumption involving a variable $p$ to generate an infinite sequence of consequences that no longer contain $p$. These consequences describe what can be obtained by successively repeating a modal step. A uniform post-interpolant would have to be a $p$-free consequence of the original formula and, at the same time, imply every member of this sequence. Thus, to prove failure of uniform interpolation, we must construct such a sequence and show that no single $p$-free consequence of the original formula can imply all its members. This leads us to study the finite approximations produced by successive iterations.

Our post-interpolation counterexample and its approximations are closely related to the fixed-point construction in the proof of Kowalski and Metcalfe's Theorem~3.1~\cite{KowalskiMetcalfe2018}; see also~\cite{KowalskiMetcalfe2019}. We draw on this intuition to establish the required finite approximation property directly by transforming cut-free proofs in Krupski and Yatmanov's calculus.

This intuition leads us to explicit counterexamples to both uniform post-interpolation and uniform pre-interpolation in every logic $L$ with $\mathrm{IEL}^{-}\subseteq L\subseteq\mathrm{IEL}$. The construction starts from a simple observation: co-reflection already provides $p\to\Box p$, so adding $\Box p\to p$ as an assumption makes $p$ behave as a fixed point of the modality. To place this fixed point between a starting condition and an observable conclusion, fix distinct variables $p,q,r$ and consider
\[
((q\to p)\land(\Box p\to p))\land(p\to r).
\]
We seek to eliminate $p$. The implication $q\to p$ provides the starting point; $\Box p\to p$ allows a boxed occurrence of $p$ to be used as $p$ again; and $p\to r$ turns information obtained about $p$ into a conclusion in the remaining vocabulary.

The proof has two main steps. First, we establish a \emph{finite approximation theorem}: in $\mathrm{IEL}$, every $p$-free consequence of the displayed formula is implied by $\Box^nq\to r$ for some $n\geq0$, with $\Box^0q=q$. Each such approximation is already a consequence of the formula in $\mathrm{IEL}^{-}$. Second, we prove \emph{strictness}: each successive approximation is strictly stronger than its predecessor, even in $\mathrm{IEL}$. Any uniform post-interpolant would therefore be implied by one approximation while having to imply its strictly stronger successor, contradicting strictness.

The same version of finite approximation yields the corresponding failure of uniform pre-interpolation for
\[
\Bigl(((q\to p)\land(\Box p\to p))\land(p\to r)\Bigr)\to r,
\]
using the strictly weakening sequence $(\Box^nq\to r)\to r$. The same arguments apply throughout the interval $\mathrm{IEL}^{-}\subseteq L\subseteq\mathrm{IEL}$.

The rest of the paper is organised as follows. Section~2 provides the definitions and proof-theoretic background. Section~3 proves the finite approximation theorem, and Section~4 establishes strictness of the approximations. Section~5 combines these results to obtain the failures of uniform post-interpolation and uniform pre-interpolation. Section~6 concludes with further questions.

\section{Preliminaries}

In this section, we recall the notions of uniform post- and pre-interpolation and the Hilbert systems for $\mathrm{IEL}^{-}$ and $\mathrm{IEL}$. We also introduce the sequent calculus $\mathsf G$ for $\mathrm{IEL}$ and establish the basic admissibility results used in the subsequent proofs.

Let $\mathsf{Prop}$ be a countably infinite set of propositional variables. The set $\mathsf{Fm}$ of formulas of the modal language $\mathcal{L}_{\Box}$ is generated by the following grammar:
\[
\varphi ::= p \mid \bot \mid (\varphi\land\varphi) \mid (\varphi\lor\varphi) \mid (\varphi\to\varphi) \mid \Box\varphi,
\]
where $p\in\mathsf{Prop}$. We write $\neg\varphi$ for $\varphi\to\bot$ and $\top$ for $\bot\to\bot$. Let $\Box^0\varphi=\varphi$ and $\Box^{n+1}\varphi=\Box(\Box^n\varphi)$ for $n\geq 0$. For a formula $\varphi$, let $\operatorname{Voc}(\varphi)$ denote the set of propositional variables occurring in $\varphi$. A formula $\varphi$ is $p$-free if $p\notin\operatorname{Voc}(\varphi)$, and a multiset is $p$-free if each of its members is $p$-free. Let $L$ be an intuitionistic modal logic in the above language. We use $\vdash_L\varphi$ to denote that $\varphi$ is provable in the Hilbert system for $L$. For a finite multiset $\Gamma$, the notation $\Gamma\vdash_L\varphi$ abbreviates $\vdash_L(\bigwedge\Gamma)\to\varphi$, with $\bigwedge\varnothing=\top$. For $\Delta=\varphi_1,\ldots,\varphi_k$, the multiset $\Box\Delta$ is $\Box\varphi_1,\ldots,\Box\varphi_k$.

\subsection{Uniform Interpolation}

We first give the definition of uniform interpolants and the uniform interpolation property. Fix a logic \(L\), a formula \(\varphi\), and a propositional variable \(p\).

\begin{definition}[Uniform post-interpolant]
\label{def:1-1}

A formula \(\chi\) is a \textbf{uniform post-interpolant} for \(\varphi\) with respect to \(p\) if:

\begin{enumerate}

\tightlist
\item \(\operatorname{Voc}(\chi)\subseteq\operatorname{Voc}(\varphi)\setminus\{p\}\);
\item \(\vdash_L\varphi\to\chi\);
\item for every \(p\)-free formula \(\psi\), if \(\vdash_L\varphi\to\psi\), then \(\vdash_L\chi\to\psi\).
\end{enumerate}
\end{definition}

\begin{definition}[Uniform pre-interpolant]
\label{def:1-2}

A formula \(\theta\) is a \textbf{uniform pre-interpolant} for \(\varphi\) with respect to \(p\) if:

\begin{enumerate}

\tightlist
\item \(\operatorname{Voc}(\theta)\subseteq\operatorname{Voc}(\varphi)\setminus\{p\}\);
\item \(\vdash_L\theta\to\varphi\);
\item for every \(p\)-free formula \(\psi\), if \(\vdash_L\psi\to\varphi\), then \(\vdash_L\psi\to\theta\).
\end{enumerate}

\end{definition}

\begin{definition}[Uniform interpolation property]
\label{def:1-3}

A logic \(L\) has the \textbf{uniform interpolation property} if uniform post-interpolants and uniform pre-interpolants exist for every formula \(\varphi\) and every propositional variable \(p\).

\end{definition}

In Definitions \ref{def:1-1} and \ref{def:1-2}, condition (1) requires the interpolant to contain neither \(p\) nor any variable absent from \(\varphi\). In condition (3), the only restriction on \(\psi\) is that it should be \(p\)-free.

\subsection{Intuitionistic Epistemic Logic}

We use the Hilbert-style systems of $\mathrm{IEL}^-$ and $\mathrm{IEL}$ introduced by Artemov and Protopopescu. The logic $\mathrm{IEL}$ properly extends $\mathrm{IEL}^{-}$ \cite{ArtemovProtopopescu2016}.

The axiom schemata of \textbf{IEL\(^{-}\)} are:

\begin{itemize}
\item the axiom schemata of intuitionistic propositional logic IPC, allowing arbitrary formulas of $\mathcal{L}_{\Box}$ as instances;
\item $(K):$   \(\Box(\varphi\to \psi)\to(\Box \varphi\to\Box \psi)\);
\item $(\text{Co-reflection}):$  \(\varphi\to\Box \varphi\).
\end{itemize}

The only inference rule is modus ponens:

\[
\frac{\varphi\qquad \varphi\to \psi}{\psi}\;\mathrm{MP}.
\]

The system \textbf{IEL} has the same axioms and rule, together with the additional axiom schema of \textbf{weak factivity}:
\[
\Box \varphi\to\neg\neg \varphi.
\]

For $L\in\{\mathrm{IEL}^{-},\mathrm{IEL}\}$, the Hilbert system satisfies the deduction theorem:
\[
\Gamma,\varphi\vdash_L\psi
\quad\Longleftrightarrow\quad
\Gamma\vdash_L\varphi\to\psi.
\]
The deduction theorem is stated in Artemov and Protopopescu~\cite[Proposition~3.3(2)]{ArtemovProtopopescu2016}.

\begin{lemma}\label{lem:box-iteration}
For all formulas $\alpha,\beta$ of $\mathcal{L}_\Box$ and integers $0\leq i\leq j$,
\[
\begin{gathered}
\vdash_{\mathrm{IEL}^{-}}\Box^i\alpha\to\Box^j\alpha,\\
\vdash_{\mathrm{IEL}^{-}}
  (\Box^j\alpha\to\beta)\to(\Box^i\alpha\to\beta).
\end{gathered}
\]
\end{lemma}

\begin{proof}
The first claim follows by iterating co-reflection and using transitivity of implication; when $i=j$, it is an instance of $\varphi\to\varphi$. The second follows from the first by modus ponens and the following instance of a theorem of intuitionistic propositional logic:
\[
(\Box^i\alpha\to\Box^j\alpha)\to
\bigl((\Box^j\alpha\to\beta)\to(\Box^i\alpha\to\beta)\bigr).
\]
\end{proof}

We use the sequent calculus of $\mathrm{IEL}$ from Krupski and Yatmanov~\cite{KrupskiYatmanov2016}, denoted by $\mathsf G$. Sequents have the form $\Gamma\Rightarrow\chi$, where $\Gamma$, the antecedent, consists of a finite and possibly empty multiset of formulas of $\mathcal{L}_{\Box}$ and $\chi$, the succedent, consists of exactly one formula of $\mathcal{L}_{\Box}$. The rules of $\mathsf G$ are given in Figure \ref{fig:iel-calculus}. We do not require a sequent calculus for $\mathrm{IEL}^{-}$ here, but we conjecture it could be obtained from $\mathsf G$ by omitting $(U)$ and adding the initial sequents $\Gamma,\bot\Rightarrow\chi$.

\begin{figure}[tbp]
\centering
\setlength{\fboxsep}{10pt}
\fbox{%
\begin{minipage}{\dimexpr\linewidth-2\fboxsep-2\fboxrule\relax}
\centering
\small
\[
\begin{gathered}
\overline{\Gamma,a\Rightarrow a}
\;(\mathrm{Id})
\qquad (a\in\mathsf{Prop}\cup\{\bot\})
\\[2ex]
\frac{\Gamma,\alpha,\beta\Rightarrow\chi}
     {\Gamma,\alpha\land\beta\Rightarrow\chi}
\;(\land L)
\qquad
\frac{\Gamma\Rightarrow\alpha
      \quad\Gamma\Rightarrow\beta}
     {\Gamma\Rightarrow\alpha\land\beta}
\;(\land R)
\\[2ex]
\frac{\Gamma,\alpha\Rightarrow\chi
      \quad\Gamma,\beta\Rightarrow\chi}
     {\Gamma,\alpha\lor\beta\Rightarrow\chi}
\;(\lor L)
\\[2ex]
\frac{\Gamma\Rightarrow\alpha_i}
     {\Gamma\Rightarrow\alpha_1\lor\alpha_2}
\;(\lor R_i)
\quad i\in\{1,2\}
\\[2ex]
\frac{\Gamma,\alpha\to\beta\Rightarrow\alpha
      \quad\Gamma,\beta\Rightarrow\chi}
     {\Gamma,\alpha\to\beta\Rightarrow\chi}
\;(\to L)
\\[2ex]
\frac{\Gamma,\alpha\Rightarrow\beta}
     {\Gamma\Rightarrow\alpha\to\beta}
\;(\to R)
\\[2ex]
\frac{\Pi,\Box\Delta,\Delta\Rightarrow\alpha}
     {\Pi,\Box\Delta\Rightarrow\Box\alpha}
\;(R_\Box)
\qquad (\text{no formula of the form }\Box \beta \text{ in } \Pi)
\\[2ex]
\frac{\Gamma\Rightarrow\Box\bot}
     {\Gamma\Rightarrow\chi}
\;(U)
\end{gathered}
\]
\end{minipage}%
}
\caption{Rules of the sequent calculus $\mathsf G$ for $\mathrm{IEL}$.}
\label{fig:iel-calculus}
\end{figure}

A \(\mathsf G\) proof is a finite tree using only the rules in Figure \ref{fig:iel-calculus}. We write $\mathsf G\vdash\Gamma\Rightarrow\varphi$ to mean that the sequent $\Gamma\Rightarrow\varphi$ has a $\mathsf G$ proof. Initial sequents have height \(0\); every other proof has height one plus the maximum height of its immediate subproofs. The principal formula of a propositional rule is the displayed formula introduced or decomposed in its conclusion. In $(R_\Box)$, the principal formula occurrences in the conclusion are all members of $\Box\Delta$ together with the succedent $\Box\alpha$. In $(U)$, the principal formula is the succedent $\chi$ of its conclusion. We say a sequent is derivable in \(\mathsf G\)  if there is a \(\mathsf G\)-proof of it.

A rule is \textbf{admissible} if derivability of its premises implies derivability of its conclusion. Such a transformation is \textbf{height-preserving} if the height of the resulting derivation does not exceed the maximum height of the premise derivations.

Krupski and Yatmanov use $\Box \bot \to \bot$ as the additional axiom of IEL, rather than $\Box \varphi \to \neg \neg \varphi$. Over IEL\(^{-}\) they are equivalent. Weak factivity with \(\varphi=\bot\), followed by the IPC theorem \(\neg\neg\bot\to\bot\), gives \(\Box\bot\to\bot\). Conversely, under the assumptions \(\Box \varphi,\varphi\to\bot\), co-reflection and distribution yield \(\Box\bot\). By $\Box\bot\to\bot$, we have \(\bot\). Applying the deduction theorem twice gives \(\Box \varphi\to\neg\neg \varphi\).

\begin{theorem}[Krupski–Yatmanov]
\label{thm:2-1}

In \(\mathsf G\), weakening and contraction rules
\[
\frac{\Gamma\Rightarrow \chi}{\Gamma,\gamma\Rightarrow \chi}\;W,
\qquad
\frac{\Gamma,\gamma,\gamma\Rightarrow \chi}{\Gamma,\gamma\Rightarrow \chi}\;\mathrm{Ctr}
\]

are height-preserving admissible. The rule
\[
\frac{\Gamma\Rightarrow \gamma\quad\Lambda,\gamma\Rightarrow \chi}
{\Gamma,\Lambda\Rightarrow \chi}\;\mathrm{Cut_m}
\]

is admissible. Moreover,

\[
\mathsf G\vdash\Gamma\Rightarrow \chi
\quad\Longleftrightarrow\quad
\vdash_{\mathrm{IEL}}(\bigwedge\Gamma)\to \chi.
\]

\end{theorem}

In the following, we sometimes use the admissible additive form of Cut:
\[
\frac{
  \Lambda\Rightarrow\varphi
  \qquad
  \Lambda,\varphi\Rightarrow\psi
}{
  \Lambda\Rightarrow\psi
}\;\mathrm{Cut}.
\]
Its admissibility follows from the admissible $\mathrm{Cut_m}$ of Theorem \ref{thm:2-1} and contraction.

Height-preserving admissibility of weakening and contraction is established in Section~4 of ~\cite{KrupskiYatmanov2016}; admissibility of Cut and the correspondence between sequent calculus and the Hilbert-style system are established in Section~6 of ~\cite{KrupskiYatmanov2016}.

Next, we claim and prove some useful facts about $\mathsf G$.

\begin{lemma}[General Identity]
\label{lem:2-2}
For every formula $\alpha$ and finite multiset $\Gamma$ of formulas, the sequent $\Gamma,\alpha\Rightarrow\alpha$ is derivable in $\mathsf G$.
\end{lemma}

\begin{proof}
We prove this claim by induction on the structure of $\alpha$. If $\alpha$ is a propositional variable or $\bot$, the sequent is initial.

\noindent For the remaining cases, we construct the following proofs. Here, ``IH'' is used for ``inductive hypothesis''.

\smallskip\noindent
\emph{Case $\alpha=\beta\land\gamma$.}
\[
\vlderivation{
  \vlin{}{\land L}
    {\Gamma,\beta\land\gamma\Rightarrow\beta\land\gamma}{
    \vliin{}{\land R}
      {\Gamma,\beta,\gamma\Rightarrow\beta\land\gamma}
      {\IHproof{\Gamma,\beta,\gamma\Rightarrow\beta}}
      {\IHproof{\Gamma,\beta,\gamma\Rightarrow\gamma}}
  }
}
\]

\smallskip\noindent
\emph{Case $\alpha=\beta\lor\gamma$.}
\[
\vlderivation{
  \vliin{}{\lor L}
    {\Gamma,\beta\lor\gamma\Rightarrow\beta\lor\gamma}
    {\vlin{}{\lor R_1}{\Gamma,\beta\Rightarrow\beta\lor\gamma}
      {\IHproof{\Gamma,\beta\Rightarrow\beta}}}
    {\vlin{}{\lor R_2}{\Gamma,\gamma\Rightarrow\beta\lor\gamma}
      {\IHproof{\Gamma,\gamma\Rightarrow\gamma}}}
}
\]

\smallskip\noindent
\emph{Case $\alpha=\beta\to\gamma$.}
\[
\vlderivation{
  \vlin{}{\to R}
    {\Gamma,\beta\to\gamma\Rightarrow\beta\to\gamma}{
    \vliin{}{\to L}
      {\Gamma,\beta,\beta\to\gamma\Rightarrow\gamma}
      {\IHproof{\Gamma,\beta,\beta\to\gamma\Rightarrow\beta}}
      {\IHproof{\Gamma,\beta,\gamma\Rightarrow\gamma}}
  }
}
\]

\smallskip\noindent
\emph{Case $\alpha=\Box\beta$.}
Write $\Gamma=\Pi,\Box\Delta$, where $\Pi$ contains no formula whose principal connective is $\Box$. Then
\[
\vlderivation{
  \vlin{}{R_\Box}{\Pi,\Box\Delta,\Box\beta\Rightarrow\Box\beta}
    {\IHproof{\Pi,\Box\Delta,\Box\beta,\Delta,\beta
      \Rightarrow\beta}}
}
\]
is a proof of $\Gamma,\Box\beta\Rightarrow\Box\beta$.

This completes the induction.
\end{proof}

\begin{lemma}[Necessitation]\label{lem:nec}
The following rule is admissible in $\mathsf G$:
\[
\vlderivation{
  \vlin{}{\mathrm{nec}}
    {\Gamma\Rightarrow\Box\alpha}
    {\vlhy{\Gamma\Rightarrow\alpha}}
}
\]
\end{lemma}

\begin{proof}
Suppose $\Gamma\Rightarrow\alpha$ is derivable. Write $\Gamma=\Pi,\Box\Delta$, where $\Pi$ contains no formula of the form $\Box\beta$. Then we have 
\[
\vlderivation{
  \vlin{}{R_\Box}
    {\Pi,\Box\Delta\Rightarrow\Box\alpha}{
    \vlin{}{W^*}
      {\Pi,\Box\Delta,\Delta\Rightarrow\alpha}
      {\vlhy{\Pi,\Box\Delta\Rightarrow\alpha}}
  }
}
\]
where $W^*$ abbreviates finitely-many applications of the admissible weakening rule. This exactly gives us $\Gamma \Rightarrow \Box \alpha$. 
\end{proof}

\begin{lemma}\label{lem:2-3}
For every formula $\chi$ and finite multiset $\Gamma$ of formulas, the following rule is admissible in $\mathsf G$:
\[
\vlderivation{
  \vlin{}{\mathrm{EFQ}}
    {\Gamma\Rightarrow\chi}
    {\vlhy{\Gamma\Rightarrow\bot}}
}
\]
In particular, $\Gamma,\bot\Rightarrow\chi$ is derivable.
\end{lemma}

\begin{proof}
Using Lemma \ref{lem:nec}, we obtain
\[
\vlderivation{
  \vlin{}{U}
    {\Gamma\Rightarrow\chi}{
    \vlin{}{\mathrm{nec}}
      {\Gamma\Rightarrow\Box\bot}
      {\vlhy{\Gamma\Rightarrow\bot}}
  }
}
\qquad
\vlderivation{
  \vlin{}{U}
    {\Gamma,\bot\Rightarrow\chi}{
    \vlin{}{\mathrm{nec}}
      {\Gamma,\bot\Rightarrow\Box\bot}{
      \vlin{}{\mathrm{Id}}
        {\Gamma,\bot\Rightarrow\bot}
        {\vlhy{}}
    }
  }
}
\]
Thus, both assertions follow without adding an initial $\bot L$ rule.
\end{proof}

We next introduce uniform substitutions and show that they preserve derivability in $\mathsf G$.

\begin{definition}[Uniform substitution]
\label{def:substitution}
A \emph{uniform substitution} is a function $\sigma\colon\mathsf{Prop}\to\mathsf{Fm}$. It extends uniquely to formulas. Writing $\alpha\sigma$ for the image of $\alpha$, we define inductively:
\[
\begin{aligned}
p\sigma &= \sigma(p),\\
\bot\sigma &= \bot,\\
(\alpha\land\beta)\sigma
  &= (\alpha\sigma)\land(\beta\sigma),\\
(\alpha\lor\beta)\sigma
  &= (\alpha\sigma)\lor(\beta\sigma),\\
(\alpha\to\beta)\sigma
  &= (\alpha\sigma)\to(\beta\sigma),\\
(\Box\alpha)\sigma &= \Box(\alpha\sigma),
\end{aligned}
\]
for every $p\in\mathsf{Prop}$ and $\alpha,\beta\in\mathsf{Fm}$. For finite multisets, let
\[
\varnothing\sigma=\varnothing,
\qquad
(\Gamma,\alpha)\sigma=\Gamma\sigma,\alpha\sigma.
\]
Finally, we write
\[
(\Gamma\Rightarrow\chi)\sigma
  = (\Gamma\sigma\Rightarrow\chi\sigma).
\]
\end{definition}

\begin{lemma}
\label{lem:2-4}
Every uniform substitution \(\sigma\) preserves derivability in \(\mathsf G\). That is, for every uniform substitution $\sigma$, the following rule is admissible in $\mathsf G$:
\[
\vlderivation{
  \vlin{}{\mathrm{Sub}}
    {\Gamma\sigma\Rightarrow\chi\sigma}
    {\vlhy{\Gamma\Rightarrow\chi}}
}
\]
\end{lemma}

\begin{proof}
We proceed by induction on the height of the given derivation. For an initial sequent $\Gamma,a\Rightarrow a$, its substituted instance $\Gamma\sigma,a\sigma\Rightarrow a\sigma$ is derivable by Lemma \ref{lem:2-2}. For propositional rules and $U$, apply the inductive hypotheses to the premises and reapply the same rule.

The only non-trivial case is that the last rule applied is  $R_\Box$. Suppose the last step is
\[
\vlderivation{
  \vlin{}{R_\Box}
    {\Pi,\Box\Delta\Rightarrow\Box\alpha}
    {\vlhy{\Pi,\Box\Delta,\Delta\Rightarrow\alpha}}
}
\]
Although $\Pi$ contains no formula whose principal connective is $\Box$, $\Pi\sigma$ may do so. Thus, decompose $\Pi\sigma=\Pi_0,\Box\Sigma$, where $\Pi_0$ contains no formula whose principal connective is $\Box$. We then have
\[
\vlderivation{
  \vlin{}{R_\Box}
    {\Pi_0,\Box\Sigma,\Box(\Delta\sigma)
      \Rightarrow\Box(\alpha\sigma)}{
    \vlin{}{W^*}
      {\Pi_0,\Box\Sigma,\Box(\Delta\sigma),\Sigma,\Delta\sigma
        \Rightarrow\alpha\sigma}
      {\IHproof{
        \Pi_0,\Box\Sigma,\Box(\Delta\sigma),\Delta\sigma
        \Rightarrow\alpha\sigma}}
  }
}
\]
Here $W^*$ denotes finitely-many applications of the admissible weakening rule, adding all members of $\Sigma$. The conclusion is exactly the substituted conclusion we want.
\end{proof}

\newpage

\section{Finite Approximation}\label{sec:3}

We establish a finite approximation theorem for the formulas introduced below, using several auxiliary admissible rules. Together with the strict implications of the next section, this theorem will yield the failure of uniform interpolation.

We first define the approximating formulas and contexts we use in the following sections.

\begin{definition}[Approximating formulas and contexts]
\label{def:approximations}
Fix pairwise distinct propositional variables $p,q,r$ for the rest of this paper. We consider the following two families of formulas, indexed by integers $n\geq 0$:
\[
\Box^nq\to r
\qquad\text{and}\qquad
(\Box^nq\to r)\to r.
\]
Let $\Omega$ be any finite multiset whose members belong to the set 
\[
\{q\to p,\;\Box p\to p,\;p\to r\}.
\]
The multiset $\Omega$ may be empty, contain repetitions, or omit any of these formulas.
\end{definition}

\subsection{Additional Admissible Rules}\label{sec:3-1}

We establish the admissibility of several auxiliary rules in $\mathsf G$ that will be used in the proof of the finite approximation theorem.

\begin{lemma}\label{lem:3-1}
Let $n,m$ be integers with $0 \leq n\leq m$. The following sequents are derivable in $\mathsf G$:
\[
\begin{gathered}
\Box^nq\Rightarrow\Box^mq,\\
(\Box^mq\to r)\Rightarrow(\Box^nq\to r).
\end{gathered}
\]
For every formula $\alpha$ and finite multiset $\Gamma$ of formulas, the following rules are admissible in $\mathsf G$:
\[
\begin{gathered}
\vlderivation{
  \vlin{}{\mathrm{Lift}_L}
    {(\Box^mq\to r),\Gamma\Rightarrow\alpha}
    {\vlhy{(\Box^nq\to r),\Gamma\Rightarrow\alpha}}
}
\\[2ex]
\vlderivation{
  \vlin{}{\mathrm{Lift}_0}
    {(\Box^mq\to r),\Gamma\Rightarrow\Box^mq}
    {\vlhy{(\Box^nq\to r),\Gamma\Rightarrow\Box^nq}}
}
\\[2ex]
\vlderivation{
  \vlin{}{\mathrm{Lift}_1}
    {(\Box^mq\to r),\Gamma\Rightarrow\Box^{m+1}q}
    {\vlhy{(\Box^nq\to r),\Gamma\Rightarrow\Box^{n+1}q}}
}
\end{gathered}
\]
\end{lemma}

\begin{proof}
For the sequent $\Box^nq\Rightarrow\Box^mq$,  apply $\mathrm{nec}$ from Lemma \ref{lem:nec} exactly $m-n$ times to the general identity $\Box^nq\Rightarrow\Box^nq$ supplied by Lemma \ref{lem:2-2}.

For the sequent $(\Box^mq\to r)\Rightarrow(\Box^nq\to r)$, using $\Box^nq\Rightarrow\Box^mq$ and the weakening rule, we have
\[
\vlderivation{
  \vlin{}{\to R}
    {(\Box^mq\to r)\Rightarrow(\Box^nq\to r)}{
    \vliin{}{\to L}
      {(\Box^mq\to r),\Box^nq\Rightarrow r}
      {\vlin{}{W}
        {(\Box^mq\to r),\Box^nq\Rightarrow\Box^mq}
        {\vlhy{\Box^nq\Rightarrow\Box^mq}}}
      {\vlin{}{\mathrm{Id}}
        {\Box^nq,r\Rightarrow r}
        {\vlhy{}}}
  }
}
\]

Using the sequent $(\Box^mq\to r)\Rightarrow(\Box^nq\to r)$, the admissibility of $\mathrm{Lift}_L$ follows by
\[
\vlderivation{
  \vliin{}{\mathrm{Cut_m}}
    {(\Box^mq\to r),\Gamma\Rightarrow\alpha}
    {\vlhy{(\Box^mq\to r)\Rightarrow(\Box^nq\to r)}}
    {\vlhy{(\Box^nq\to r),\Gamma\Rightarrow\alpha}}
}
\]
Here, $\alpha$ and $\Gamma$ are arbitrary. In particular, $\alpha$ here need not be $p$-free. 

For $k\in\{0,1\}$, the admissibility of $\mathrm{Lift}_k$ follows from the following derivation, using $\mathrm{Lift}_L$ and the sequent $\Box^{n+k} q \Rightarrow \Box^{m+k}q$:
\[
\vlderivation{
  \vliin{}{\mathrm{Cut_m}}
    {(\Box^mq\to r),\Gamma\Rightarrow\Box^{m+k}q}
    {\vlin{}{\mathrm{Lift}_L}
      {(\Box^mq\to r),\Gamma\Rightarrow\Box^{n+k}q}
      {\vlhy{(\Box^nq\to r),\Gamma\Rightarrow\Box^{n+k}q}}}
    {\vlhy{\Box^{n+k}q\Rightarrow\Box^{m+k}q}}
}
\]
\end{proof}

Notice that, by one more application of $\to R$, we can get $\mathsf G \vdash \Rightarrow \Box^n q \to \Box^m q$ and $\mathsf G \vdash \Rightarrow (\Box^m q \to r) \to (\Box^n q \to r)$.

\begin{lemma}\label{lem:3-2}
Let $\Gamma$ be a finite multiset of $p$-free formulas, let $\alpha$ be a $p$-free formula, and let $\Omega$ be as in Definition \ref{def:approximations}. Then the following rule is admissible in $\mathsf G$:
\[
\vlderivation{
  \vlin{}{\mathrm{Elim}_p}
    {\Gamma,r\Rightarrow\alpha}
    {\vlhy{\Omega,\Gamma,p\Rightarrow\alpha}}
}
\]
\end{lemma}

\begin{proof}
Suppose that $\Omega,\Gamma,p\Rightarrow\alpha$ is derivable. Let $\sigma$ be the uniform substitution defined by $\sigma(p)=r$ and $\sigma(s)=s$ for every propositional variable $s\neq p$. Since $\Gamma$ and $\alpha$ are $p$-free, Lemma \ref{lem:2-4} yields $\Omega\sigma,\Gamma,r\Rightarrow\alpha$. 

Every formula in $\Omega\sigma$ is one of $q\to r$, $\Box r\to r$, and $r\to r$. We show how to eliminate these formulas from the antecedent.

Suppose, for example, that a designated occurrence of $q\to r$ is present, so that the sequent has the form \[
\Omega',\Gamma,r, q\to r \Rightarrow\alpha,
\]
\noindent where $\Omega\sigma=\Omega',q\to r$.

Then we obtain the following derivation:
\[
\vlderivation{
  \vliin{}{\mathrm{Cut}}
    {\Omega',\Gamma,r\Rightarrow\alpha}
    {\vlin{}{\to R}
      {\Omega',\Gamma,r\Rightarrow q\to r}
      {\vlin{}{\mathrm{Id}}
        {\Omega',\Gamma,r,q\Rightarrow r}
        {\vlhy{}}}}
    {\vlhy{\Omega',\Gamma,r,q\to r\Rightarrow\alpha}}
}
\]
Thus, one designated occurrence of $q\to r$ has been eliminated. The other two cases are similar. Repeating this procedure for all occurrences in the finite multiset $\Omega\sigma$, we obtain a derivation of $\Gamma,r\Rightarrow\alpha$.
\end{proof}

\subsection{Finite Approximation Theorem}
Using the admissible rules established above, we show that in any derivation of $\Omega,\Gamma\Rightarrow\alpha$ with $\Gamma$ and $\alpha$ $p$-free, the context $\Omega$ can be replaced by $\Box^nq\to r$ for some $n\geq0$. To carry out the induction, we prove this together with two auxiliary assertions concerning the succedents $p$ and $\Box p$.

\begin{theorem}[Finite approximation]
\label{thm:3-3}
Let $\Omega$ be as in Definition \ref{def:approximations}, let $\Gamma$ be a finite multiset of $p$-free formulas, and let $\alpha$ be a $p$-free formula. The following rules are admissible in $\mathsf G$ with an existential choice of $n$: whenever a premise is derivable in $\mathsf G$, there exists an integer $n\geq0$ such that the corresponding conclusion is derivable in $\mathsf G$.
\[
\begin{gathered}
\vlderivation{
  \vlin{}{\mathrm{P}_0}
    {(\Box^nq\to r),\Gamma\Rightarrow\alpha}
    {\vlhy{\Omega,\Gamma\Rightarrow\alpha}}
}
\\[2ex]
\vlderivation{
  \vlin{}{\mathrm{P}_1}
    {(\Box^nq\to r),\Gamma\Rightarrow\Box^nq}
    {\vlhy{\Omega,\Gamma\Rightarrow p}}
}
\\[2ex]
\vlderivation{
  \vlin{}{\mathrm{P}_2}
    {(\Box^nq\to r),\Gamma\Rightarrow\Box^{n+1}q}
    {\vlhy{\Omega,\Gamma\Rightarrow\Box p}}
}
\end{gathered}
\]
\end{theorem}

\begin{proof}
We prove the admissibility of $(\mathrm P_0)$, $(\mathrm P_1)$, and $(\mathrm P_2)$ by simultaneous induction on the height of a derivation of each of their premises. Fix a derivation of $\Omega,\Gamma\Rightarrow\zeta$, where $\zeta$ is a $p$-free formula $\alpha$, $p$, or $\Box p$. For every integer $k\geq0$, put
\[
\tau_k(\zeta)=
\begin{cases}
\alpha, & \text{if $\zeta=\alpha$ is $p$-free},\\
\Box^kq, & \text{if $\zeta=p$},\\
\Box^{k+1}q, & \text{if $\zeta=\Box p$}.
\end{cases}
\]
Our task is to derive $(\Box^nq\to r),\Gamma\Rightarrow\tau_n(\zeta)$ for some $n\geq0$. We distinguish cases according to the last rule of the derivation.

\smallskip\noindent
\textbf{Case 1: the sequent is initial.}

If $\Omega,\Gamma\Rightarrow a$ is initial, then $a$ is a propositional variable or $\bot$, and $a\in\Gamma$, since every member of $\Omega$ is an implication. Thus $a$ is $p$-free. Hence, only $(\mathrm{P_0})$ is possible in this case. Choose $n=0$. The sequent \((q\to r),\Gamma\Rightarrow a\) is initial as well.

\smallskip\noindent
\textbf{Case 2: the last rule-application is $\land L$.}

Since every formula in \(\Omega\) has implication as its principal connective, any principal formula of \(\land L\) in the conclusion must belong to the \(p\)-free \(\Gamma\).

Write $\Gamma=\Gamma_0,\beta\land\gamma$, where $\beta$ and $\gamma$ are $p$-free. The last step is
\[
\vlderivation{
  \vlin{}{\land L}
    {\Omega,\Gamma_0,\beta\land\gamma\Rightarrow\zeta}
    {\vlhy{\Omega,\Gamma_0,\beta,\gamma\Rightarrow\zeta}}
}
\]
Apply the inductive hypothesis for $(\mathrm P_i)$ to its premise, where $i\in \{0,1,2\}$ according as $\zeta$ is $p$-free, $p$, or $\Box p$. Let $n$ be the index supplied by this inductive hypothesis. Using the same $n$ for the conclusion, we obtain
\[
\vlderivation{
  \vlin{}{\land L}
    {(\Box^nq\to r),\Gamma_0,\beta\land\gamma
      \Rightarrow\tau_n(\zeta)}
    {\IHcase{i}{
      (\Box^{n}q\to r),\Gamma_0,\beta,\gamma
      \Rightarrow\tau_{n}(\zeta)}}
}
\]
\noindent where $\mathrm{IH}_i$ represents the derivation supplied by the inductive hypothesis for $\mathrm{P}_i$.

Hence, all three admissible rules have been proved in this case.

\smallskip\noindent
\textbf{Case 3: the last rule-application is $\lor L$.}

Since every formula in $\Omega$ is an implication, the principal formula of $\lor L$ must belong to the $p$-free $\Gamma$. Write $\Gamma=\Gamma_0,\beta\lor\gamma$, where $\beta$ and $\gamma$ are $p$-free. The last step is
\[
\vlderivation{
  \vliin{}{\lor L}
    {\Omega,\Gamma_0,\beta\lor\gamma\Rightarrow\zeta}
    {\vlhy{\Omega,\Gamma_0,\beta\Rightarrow\zeta}}
    {\vlhy{\Omega,\Gamma_0,\gamma\Rightarrow\zeta}}
}
\]
Applying the inductive hypothesis for $(\mathrm P_i)$ to each premise, where $i=0,1,2$ according as $\zeta$ is $p$-free, $p$, or $\Box p$, we obtain the derivations of $(\Box^{n_1}q\to r),\Gamma_0,\beta \Rightarrow\tau_{n_1}(\zeta)$ and $(\Box^{n_2}q\to r),\Gamma_0,\gamma \Rightarrow\tau_{n_2}(\zeta)$, where $n_1$ and $n_2$ are the indices obtained from the left and right premises, respectively. Set $n=\max\{n_1,n_2\}$. Then we have
\[
\vlderivation{
  \vliin{}{\lor L}
    {(\Box^nq\to r),\Gamma_0,\beta\lor\gamma
      \Rightarrow\tau_n(\zeta)}
    {\vlin{}{\text{Lem. \ref{lem:3-1}}}
      {(\Box^nq\to r),\Gamma_0,\beta\Rightarrow\tau_n(\zeta)}
      {\IHcase{i}{
        (\Box^{n_1}q\to r),\Gamma_0,\beta
        \Rightarrow\tau_{n_1}(\zeta)}}}
    {\vlin{}{\text{Lem. \ref{lem:3-1}}}
      {(\Box^nq\to r),\Gamma_0,\gamma\Rightarrow\tau_n(\zeta)}
      {\IHcase{i}{
        (\Box^{n_2}q\to r),\Gamma_0,\gamma
        \Rightarrow\tau_{n_2}(\zeta)}}}
}
\]
For the uses of Lemma \ref{lem:3-1}, in case $(\mathrm P_0)$, where $\zeta$ is $p$-free, we use $\mathrm{Lift}_L$, only changing the antecedent formula. In cases $(\mathrm P_1)$ and $(\mathrm P_2)$, where $\zeta$ is $p$ or $\Box p$, we use $\mathrm{Lift}_0$ or $\mathrm{Lift}_1$, respectively, changing the antecedent formula and the corresponding succedent formula as well.

Hence, all three admissible rules have been proved in this case.

\textbf{Case 4: the last rule-application is $\land R$, $\lor R$, or $\to R$.} 

This case is similar to Cases 2 and 3, but simpler: only \((\mathrm P_0)\) is possible, since neither $p$ nor $\Box p$ can be the principal formula of a propositional right rule.

\emph{The last rule is $\land R$.}
Write $\zeta=\beta\land\gamma$. Both $\beta$ and $\gamma$ are $p$-free. Apply the inductive hypothesis for $(\mathrm P_0)$ to the premises $\Omega,\Gamma\Rightarrow\beta$ and $\Omega,\Gamma\Rightarrow\gamma$, obtaining indices $n_1$ and $n_2$, respectively. Set $n=\max\{n_1,n_2\}$. Using $\mathrm{Lift}_L$ on both transformed premises and reapplying $\land R$, we obtain
\[
\vlderivation{
  \vliin{}{\land R}
    {(\Box^nq\to r),\Gamma\Rightarrow\beta\land\gamma}
    {\vlin{}{\mathrm{Lift}_L}
      {(\Box^nq\to r),\Gamma\Rightarrow\beta}
      {\IHcase{0}{(\Box^{n_1}q\to r),\Gamma\Rightarrow\beta}}}
    {\vlin{}{\mathrm{Lift}_L}
      {(\Box^nq\to r),\Gamma\Rightarrow\gamma}
      {\IHcase{0}{(\Box^{n_2}q\to r),\Gamma\Rightarrow\gamma}}}
}
\]

\emph{The last rule is $\lor R_j$, with $j\in\{1,2\}$.}
Write $\zeta=\beta_1\lor\beta_2$. Since $\beta_j$ is $p$-free, applying the inductive hypothesis for $(\mathrm P_0)$ to the premise $\Omega,\Gamma\Rightarrow\beta_j$ yields an index $n$. Using the same $n$ for the conclusion, we obtain
\[
\vlderivation{
  \vlin{}{\lor R_j}
    {(\Box^nq\to r),\Gamma\Rightarrow\beta_1\lor\beta_2}
    {\IHcase{0}{(\Box^nq\to r),\Gamma\Rightarrow\beta_j}}
}
\]

\emph{The last rule is $\to R$.}
Write $\zeta=\beta\to\gamma$. Both $\beta$ and $\gamma$ are $p$-free, so the inductive hypothesis for $(\mathrm P_0)$ applies to the premise $\Omega,\Gamma,\beta\Rightarrow\gamma$ with the enlarged $p$-free context $\Gamma,\beta$. Let $n$ be the index supplied by inductive hypothesis. Using the same $n$ for the conclusion, we obtain
\[
\vlderivation{
  \vlin{}{\to R}
    {(\Box^nq\to r),\Gamma\Rightarrow\beta\to\gamma}
    {\IHcase{0}{
      (\Box^nq\to r),\Gamma,\beta\Rightarrow\gamma}}
}
\]
Hence $(\mathrm P_0)$ holds for each propositional right rule, while $(\mathrm P_1)$ and $(\mathrm P_2)$ are both impossible in this case.

\smallskip\noindent
\textbf{Case 5: the last rule-application is $R_\Box$.}

Let $\Gamma=\Pi,\Box\Delta$, where $\Pi$ contains no formula whose principal connective is $\Box$. The last step is
\[
\vlderivation{
  \vlin{}{R_\Box}
    {\Omega,\Pi,\Box\Delta\Rightarrow\Box\beta}
    {\vlhy{\Omega,\Pi,\Box\Delta,\Delta\Rightarrow\beta}}
}
\]
Notice that $(\mathrm P_1)$ is impossible because the succedent of $(\mathrm P_1)$ is $p$. Then only cases $(\mathrm P_0)$ or $(\mathrm P_2)$ are possible.

\emph{Case $(\mathrm P_0)$.}
Here $\Box\beta$ is $p$-free, so $\beta$ is $p$-free as well. Applying the inductive hypothesis for $(\mathrm P_0)$ to the premise $\Omega,\Pi,\Box\Delta,\Delta\Rightarrow\beta$ yields an index $n$. Using the same $n$ for the conclusion, we obtain
\[
\vlderivation{
  \vlin{}{R_\Box}
    {(\Box^nq\to r),\Pi,\Box\Delta\Rightarrow\Box\beta}
    {\IHcase{0}{
      (\Box^nq\to r),\Pi,\Box\Delta,\Delta
      \Rightarrow\beta}}
}
\]
as required by $(\mathrm{P}_0)$.

\emph{Case $(\mathrm P_2)$.}
Here $\zeta=\Box p$, so $\beta=p$. Applying the inductive hypothesis for $(\mathrm P_1)$ to the premise $\Omega,\Pi,\Box\Delta,\Delta\Rightarrow p$ yields an index $n$. Using the same $n$ for the conclusion, we obtain
\[
\vlderivation{
  \vlin{}{R_\Box}
    {(\Box^nq\to r),\Pi,\Box\Delta\Rightarrow\Box^{n+1}q}
    {\IHcase{1}{
      (\Box^nq\to r),\Pi,\Box\Delta,\Delta
      \Rightarrow\Box^nq}}
}
\]
as required by $(\mathrm{P}_2)$.

In both cases, the side condition of $R_\Box$ holds, because $\Pi$ contains no formula whose principal connective is $\Box$, and the additional formula $\Box^nq\to r$ is an implication.

Thus, $(\mathrm P_0)$ and $(\mathrm P_2)$ hold, while $(\mathrm{P_1})$ is impossible in this case.

\smallskip\noindent
\textbf{Case 6: the last rule-application is $U$.}

The last step is
\[
\vlderivation{
  \vlin{}{U}
    {\Omega,\Gamma\Rightarrow\zeta}
    {\vlhy{\Omega,\Gamma\Rightarrow\Box\bot}}
}
\]
Regardless of whether $\zeta$ is $\alpha$, $p$ or $\Box p$, the premise succedent $\Box\bot$ is $p$-free. Applying  the inductive hypothesis for $(\mathrm P_0)$ to the premise yields an index $n$. Using the same $n$ for the conclusion, we obtain
\[
\vlderivation{
  \vlin{}{U}
    {(\Box^nq\to r),\Gamma\Rightarrow\tau_n(\zeta)}
    {\IHcase{0}{(\Box^nq\to r),\Gamma\Rightarrow\Box\bot}}
}
\]
Since $\zeta$ may be $\alpha$, $p$, or $\Box p$, this proves $(\mathrm P_0)$, $(\mathrm P_1)$, and $(\mathrm P_2)$ in this case.

\smallskip\noindent
\textbf{Case 7: the last rule-application is $\to L$.} 
Four subcases are considered, for this is the only case in which a formula in $\Omega$ could be principal.

\smallskip\noindent
\emph{Subcase 7.1: the principal formula belongs to $\Gamma$.}
Write $\Gamma=\Gamma_0,\beta\to\gamma$, where $\beta$ and $\gamma$ are $p$-free. The last step is
\[
\vlderivation{
  \vliin{}{\to L}
    {\Omega,\Gamma_0,\beta\to\gamma\Rightarrow\zeta}
    {\vlhy{\Omega,\Gamma_0,\beta\to\gamma\Rightarrow\beta}}
    {\vlhy{\Omega,\Gamma_0,\gamma\Rightarrow\zeta}}
}
\]
Since $\beta$ is $p$-free, apply the inductive hypothesis for $(\mathrm P_0)$ to the first premise $\Omega,\Gamma_0,\beta\to\gamma\Rightarrow\beta$, obtaining an index $n_1$. Apply the inductive hypothesis for $(\mathrm P_i)$ to the second premise $\Omega,\Gamma_0,\gamma\Rightarrow\zeta$, obtaining an index $n_2$, where $i\in \{0,1,2\}$ according as $\zeta$ is $p$-free, $p$, or $\Box p$. Set $n=\max\{n_1,n_2\}$. For the first premise, use $\mathrm{Lift}_L$. For the second, use $\mathrm{Lift}_L$, $\mathrm{Lift}_0$, or $\mathrm{Lift}_1$ in cases $(\mathrm P_0)$, $(\mathrm P_1)$, or $(\mathrm P_2)$, respectively. These steps allow us to obtain
\[
\vlderivation{
  \vliin{}{\to L}
    {(\Box^nq\to r),\Gamma_0,\beta\to\gamma
      \Rightarrow\tau_n(\zeta)}
    {\vlin{}{\mathrm{Lift}_L}
      {(\Box^nq\to r),\Gamma_0,\beta\to\gamma\Rightarrow\beta}
      {\IHcase{0}{
        (\Box^{n_1}q\to r),\Gamma_0,\beta\to\gamma
        \Rightarrow\beta}}}
    {\vlin{}{\text{Lem. \ref{lem:3-1}}}
      {(\Box^nq\to r),\Gamma_0,\gamma\Rightarrow\tau_n(\zeta)}
      {\IHcase{i}{
        (\Box^{n_2}q\to r),\Gamma_0,\gamma
        \Rightarrow\tau_{n_2}(\zeta)}}}
}
\]
Hence all three admissible rules hold in this subcase.

\smallskip\noindent
\emph{Subcase 7.2: the principal formula is $q\to p$.}
Write $\Omega=\Omega',q\to p$. The last step is
\[
\vlderivation{
  \vliin{}{\to L}
    {\Omega',\Gamma,q\to p\Rightarrow\zeta}
    {\vlhy{\Omega',\Gamma,q\to p\Rightarrow q}}
    {\vlhy{\Omega',\Gamma,p\Rightarrow\zeta}}
}
\]
Since $q$ is $p$-free, apply the inductive hypothesis for $(\mathrm P_0)$ to the first premise $\Omega',\Gamma,q\to p\Rightarrow q$, obtaining an index $n$. We use this same $n$ for the conclusion in each of the three cases below. For $k\in\{n,n+1\}$, Lemma \ref{lem:3-1} provides a derivation of $q\Rightarrow\Box^kq$. Hence
\[
\vlderivation{
  \vliin{}{\mathrm{Cut_m}}
    {(\Box^nq\to r),\Gamma\Rightarrow\Box^kq}
    {\IHcase{0}{(\Box^nq\to r),\Gamma\Rightarrow q}}
    {\vlhy{q\Rightarrow\Box^kq}}
}
\tag{D1}\label{eq:approx-qp-derivation}
\]

\emph{Case $(\mathrm P_0)$.}
Here $\zeta=\alpha$ is $p$-free. Use \eqref{eq:approx-qp-derivation} with $k=n$, and apply $\mathrm{Elim}_p$ from Lemma \ref{lem:3-2} to the second premise $\Omega',\Gamma,p\Rightarrow\alpha$. Then
\[
\vlderivation{
  \vliin{}{\to L}
    {(\Box^nq\to r),\Gamma\Rightarrow\alpha}
    {\vlhy{(\Box^nq\to r),\Gamma\Rightarrow\Box^nq}}
    {\vlin{}{\mathrm{Elim}_p}
      {\Gamma,r\Rightarrow\alpha}
      {\vlhy{\Omega',\Gamma,p\Rightarrow\alpha}}}
}
\]
This proves $(\mathrm P_0)$.

\emph{Case $(\mathrm P_1)$.}
Here $\zeta=p$. Taking $k=n$ in \eqref{eq:approx-qp-derivation} gives $(\Box^nq\to r),\Gamma\Rightarrow\Box^nq$, as required.

\emph{Case $(\mathrm P_2)$.}
Here $\zeta=\Box p$. Taking $k=n+1$ in \eqref{eq:approx-qp-derivation} gives $(\Box^nq\to r),\Gamma\Rightarrow\Box^{n+1}q$, as required.

\smallskip\noindent
\emph{Subcase 7.3: the principal formula is $\Box p\to p$.}
Write $\Omega=\Omega',\Box p\to p$. The last step is
\[
\vlderivation{
  \vliin{}{\to L}
    {\Omega',\Gamma,\Box p\to p\Rightarrow\zeta}
    {\vlhy{\Omega',\Gamma,\Box p\to p\Rightarrow\Box p}}
    {\vlhy{\Omega',\Gamma,p\Rightarrow\zeta}}
}
\]
Apply the inductive hypothesis for $(\mathrm P_2)$ to the first premise $\Omega',\Gamma,\Box p\to p\Rightarrow\Box p$, obtaining an index $n'$ and a derivation of $(\Box^{n'}q\to r),\Gamma\Rightarrow\Box^{n'+1}q$.

For $(\mathrm P_0)$ and $(\mathrm P_1)$, set $n=n'+1$. Apply $\mathrm{Lift}_L$ to change the antecedent while keeping the succedent unchanged:
\[
\vlderivation{
  \vlin{}{\mathrm{Lift}_L}
    {(\Box^nq\to r),\Gamma\Rightarrow\Box^nq}
    {\IHcase{2}{
      (\Box^{n'}q\to r),\Gamma\Rightarrow\Box^{n'+1}q}}
}
\tag{D2}\label{eq:approx-boxp-lift}
\]

\emph{Case $(\mathrm P_0)$.}
Here $\zeta=\alpha$ is $p$-free. Use \eqref{eq:approx-boxp-lift}, and apply $\mathrm{Elim}_p$ to the second premise $\Omega',\Gamma,p\Rightarrow\alpha$ of $\rightarrow L$. Then $\to L$ gives
\[
\vlderivation{
  \vliin{}{\to L}
    {(\Box^nq\to r),\Gamma\Rightarrow\alpha}
    {\vlhy{(\Box^nq\to r),\Gamma\Rightarrow\Box^nq}}
    {\vlin{}{\mathrm{Elim}_p}
      {\Gamma,r\Rightarrow\alpha}
      {\vlhy{\Omega',\Gamma,p\Rightarrow\alpha}}}
}
\]
This proves $(\mathrm P_0)$ with $n=n'+1$.

\emph{Case $(\mathrm P_1)$.}
Here $\zeta=p$. Derivation \eqref{eq:approx-boxp-lift} already has the required conclusion, with $n=n'+1$.

\emph{Case $(\mathrm P_2)$.}
Here $\zeta=\Box p$. The inductive hypothesis applied to the first premise $\Omega',\Gamma,\Box p\to p\Rightarrow\Box p$ has already yielded a derivation of $(\Box^{n'}q\to r),\Gamma\Rightarrow\Box^{n'+1}q$. Taking $n=n'$ gives the required conclusion.

\smallskip\noindent
\emph{Subcase 7.4: the principal formula is $p\to r$.}
Write $\Omega=\Omega',p\to r$. The last step is
\[
\vlderivation{
  \vliin{}{\to L}
    {\Omega',\Gamma,p\to r\Rightarrow\zeta}
    {\vlhy{\Omega',\Gamma,p\to r\Rightarrow p}}
    {\vlhy{\Omega',\Gamma,r\Rightarrow\zeta}}
}
\]
Apply the inductive hypothesis for $(\mathrm P_1)$ to the first premise $\Omega',\Gamma,p\to r\Rightarrow p$, obtaining an index $n_1$. Since $r$ is $p$-free, apply the inductive hypothesis for $(\mathrm P_i)$ to the second premise $\Omega',\Gamma,r\Rightarrow\zeta$ with the enlarged $p$-free context $\Gamma,r$, obtaining an index $n_2$, where $i=0,1,2$ according as $\zeta$ is $p$-free, $p$, or $\Box p$. Set $n=\max\{n_1,n_2\}$. First, $\mathrm{Lift}_0$ and $\to L$ give
\[
\vlderivation{
  \vliin{}{\to L}
    {(\Box^nq\to r),\Gamma\Rightarrow r}
    {\vlin{}{\mathrm{Lift}_0}
      {(\Box^nq\to r),\Gamma\Rightarrow\Box^nq}
      {\IHcase{1}{
        (\Box^{n_1}q\to r),\Gamma\Rightarrow\Box^{n_1}q}}}
    {\vlin{}{\mathrm{Id}}{\Gamma,r\Rightarrow r}{\vlhy{}}}
}
\tag{D3}\label{eq:approx-pr-derivation}
\]
Raise the index of $(\Box^{n_2}q\to r),\Gamma,r\Rightarrow\tau_{n_2}(\zeta)$ from $n_2$ to $n$ using $\mathrm{Lift}_L$, $\mathrm{Lift}_0$, or $\mathrm{Lift}_1$ in cases $(\mathrm P_0)$, $(\mathrm P_1)$, or $(\mathrm P_2)$, respectively. Combining this derivation with \eqref{eq:approx-pr-derivation} by the admissible Cut rule gives us
\[
\vlderivation{
  \vliin{}{\mathrm{Cut}}
    {(\Box^nq\to r),\Gamma\Rightarrow\tau_n(\zeta)}
    {\vltr{\text{\ref{eq:approx-pr-derivation}}}
      {(\Box^nq\to r),\Gamma\Rightarrow r}
      {\vlhy{}}{\vlhy{\quad}}{\vlhy{}}}
    {\vlin{}{\text{Lem. \ref{lem:3-1}}}
      {(\Box^nq\to r),\Gamma,r\Rightarrow\tau_n(\zeta)}
      {\IHcase{i}{
        (\Box^{n_2}q\to r),\Gamma,r
        \Rightarrow\tau_{n_2}(\zeta)}}}
}
\]
For $i=0,1,2$, this gives precisely the conclusion of $(\mathrm P_0)$, $(\mathrm P_1)$, or $(\mathrm P_2)$, respectively. Hence all three admissible rules hold in this subcase.

This exhausts all the cases of the simultaneous induction.
\end{proof}

\subsection{Consequences for the Hilbert-Style Systems}
\label{sec:hilbert-consequences}

We now formulate the finite approximation result in terms of derivability in the Hilbert system for $\mathrm{IEL}$. We first establish two implications that are derivable in $\mathrm{IEL}^{-}$.
\begin{lemma}
\label{lem:4-1}

For every \(n\geq0\),

\[
\vdash_{\mathrm{IEL}^{-}}
\Bigl(((q\to p)\land(\Box p\to p))\land(p\to r)\Bigr)
\to(\Box^n q\to r),
\]

\[
\vdash_{\mathrm{IEL}^{-}}
\bigl((\Box^n q\to r)\to r\bigr)
\to\Bigl(\bigl(((q\to p)\land(\Box p\to p))\land(p\to r)\bigr)\to r\Bigr).
\]

\end{lemma}

\begin{proof}

Under the assumptions $q\to p,\Box p\to p,p\to r$, we prove $\Box^nq\to p$ by induction on $n$. For \(n=0\), this is the assumption \(q\to p\). For the induction step, suppose that $\Box^n q \to p$ has been derived for some $n\ge0$. Co-reflection gives \(\Box(\Box^n q\to p)\). By axiom $(K)$ and modus ponens, this yields \(\Box^{n+1}q\to\Box p\). With \(\Box p\to p\), by transitivity of implication, we get \(\Box^{n+1}q\to p\). Thus $\Box^nq\to p$ is derivable for every $n\geq0$. Combining it with $p\to r$ gives $\Box^nq\to r$, including $n=0$. The deduction theorem yields the first implication.

For the second implication, assume \((\Box^n q\to r)\to r\) and \(((q\to p)\land(\Box p\to p))\land(p\to r)\). The first implication supplies \(\Box^n q\to r\), and by modus ponens we have \(r\). Applying the deduction theorem twice, we get the second implication.

\end{proof}

\begin{theorem}
\label{thm:4-2}

If \(\Gamma\) and \(\alpha\) are \(p\)-free, then
\[
\begin{aligned}
&\Gamma,\bigl((q\to p)\land(\Box p\to p)\bigr)\land(p\to r)
\vdash_{\mathrm{IEL}}\alpha\\
&\qquad\Longleftrightarrow\quad
\exists n\geq0\;\Gamma,(\Box^n q\to r)\vdash_{\mathrm{IEL}}\alpha.
\end{aligned}
\]

\end{theorem}

\begin{proof}

For the forward direction, propositional reasoning replaces the conjunction in the assumptions by its three conjuncts. The Hilbert--sequent correspondence in Theorem \ref{thm:2-1} then gives a cut-free $\mathsf G$ proof of
\[
q\to p,\Box p\to p,p\to r,\Gamma\Rightarrow\alpha.
\]
Apply $(\mathrm P_0)$ of Theorem \ref{thm:3-3}. For some $n\geq0$, this yields $(\Box^nq\to r),\Gamma\Rightarrow\alpha$. The same Hilbert–sequent correspondence gives $\Gamma,(\Box^nq\to r)\vdash_{\mathrm{IEL}}\alpha$.

For the converse, it is immediate by the first implication of Lemma \ref{lem:4-1}.

\end{proof}

In particular, for every \(p\)-free formula \(\alpha\),
\[
\begin{aligned}
&\vdash_{\mathrm{IEL}}
\Bigl(((q\to p)\land(\Box p\to p))\land(p\to r)\Bigr)\to \alpha\\
&\qquad\Longleftrightarrow\quad
\exists n\geq0\;\vdash_{\mathrm{IEL}}(\Box^n q\to r)\to \alpha.
\end{aligned}
\]

\begin{corollary}
\label{cor:4-3}

For every \(p\)-free formula \(\beta\),
\[
\begin{aligned}
&\vdash_{\mathrm{IEL}}
\beta\to\Bigl(\bigl(((q\to p)\land(\Box p\to p))\land(p\to r)\bigr)\to r\Bigr)\\
&\qquad\Longleftrightarrow\quad
\exists n\geq0\;\vdash_{\mathrm{IEL}}\beta\to\bigl((\Box^n q\to r)\to r\bigr).
\end{aligned}
\]

\end{corollary}

\begin{proof}

For the forward direction, we see 

\[
\vdash_{\mathrm{IEL}} \beta\to\Bigl(\bigl(((q\to p)\land(\Box p\to p))\land(p\to r)\bigr)\to r\Bigr) 
\]
is equivalent to
\[
\beta,\bigl((q\to p)\land(\Box p\to p)\bigr)\land(p\to r)
\vdash_{\mathrm{IEL}}r.
\]

Apply Theorem \ref{thm:4-2} with \(\Gamma=\{\beta\}\) and \(\alpha=r\). This gives \(\beta,(\Box^n q\to r)\vdash_{\mathrm{IEL}}r\) for some \(n\). Applying the deduction theorem gives the required implication.

For the converse, it is immediate by the second implication of  Lemma \ref{lem:4-1}.

\end{proof}

\section{Strict Implications}

We say that a formula $\varphi$ is \emph{strictly stronger} than a formula $\psi$ in a logic $L$, or equivalently that $\psi$ is \emph{strictly weaker} than $\varphi$ in $L$, if $\vdash_L\varphi\to\psi$ but $\nvdash_L\psi\to\varphi$. In this section, we show that, in every logic $L$ with $\mathrm{IEL}^{-}\subseteq L\subseteq\mathrm{IEL}$, as $n$ increases, the formulas $\Box^nq\to r$ become strictly stronger, while $(\Box^nq\to r)\to r$ become strictly weaker. 

We first establish two non-derivability results in $\mathsf G$.

\begin{lemma}
\label{lem:5-1}

Let \(\Delta\) be a finite multiset all of whose members have the form \(\Box^iq\), with \(i\geq0\), where $q$ is the propositional variable we have fixed in Definition \ref{def:approximations}. For every \(k\geq0\),
\[
\mathsf G\nvdash\Delta\Rightarrow\Box^k\bot.
\]

\end{lemma}

\begin{proof}

We prove the following claim: for each natural number $h$, for every finite multiset $\Delta$ whose members have the form $\Box^iq$ with $i\geq0$ and every $k\geq0$, there is no $\mathsf G$ derivation of $\Delta\Rightarrow\Box^k\bot$ of height at most $h$. We prove this by induction on $h$.

\smallskip\noindent
\emph{Base case: $h=0$.}
Such a derivation would have to be initial. If $k=0$, this would require $\bot\in\Delta$, which is impossible. If $k>0$, the succedent is neither a propositional variable nor $\bot$, so the sequent cannot be initial either.

\smallskip\noindent
\emph{Inductive step.}
Assume the statement holds for some $h\geq 0$. Suppose for contradiction that, for some $\Delta$ as above and $k\geq0$, the sequent $\Delta\Rightarrow\Box^k\bot$ has a $\mathsf G$ derivation of height at most $h+1$. The inductive hypothesis states that, for every finite multiset $\Lambda$ whose members have the form $\Box^iq$ with $i\geq0$ and every $\ell\geq0$, there is no $\mathsf G$ derivation of $\Lambda\Rightarrow\Box^\ell\bot$ of height at most $h$.

\smallskip
Observe that no member of \(\Delta\) has a propositional connective as its principal connective, so no propositional left rule applies. Nor does the succedent permit a propositional right rule. Thus, the last rule is $U$ or $R_\Box$.

\smallskip
\emph{The last rule-application is $U$.} The last step is
\[
\vlderivation{
  \vlin{}{U}
    {\Delta\Rightarrow\Box^k\bot}
    {\vlhy{\Delta\Rightarrow\Box\bot}}
}
\]
The premise $\Delta\Rightarrow\Box\bot$ has a derivation of height at most $h$, contradicting the inductive hypothesis with $\Lambda=\Delta$ and $\ell=1$.

\smallskip
\emph{The last rule-application is $R_\Box$.} Then $k\geq1$. Write $\Delta=\Pi,\Box\Sigma$, where $\Pi$ contains no formula whose principal connective is $\Box$. The last step is
\[
\vlderivation{
  \vlin{}{R_\Box}
    {\Pi,\Box\Sigma\Rightarrow\Box^k\bot}
    {\vlhy{\Pi,\Box\Sigma,\Sigma
      \Rightarrow\Box^{k-1}\bot}}
}
\]
Every formula in $\Pi$ must be $q=\Box^0q$. Every formula in $\Box\Sigma$ has the form $\Box^i q$ for some $i\geq 1$. Thus, every member of $\Pi,\Box\Sigma,\Sigma$ still has the form $\Box^i q$ for some $i\geq0$. The premise $\Pi,\Box\Sigma,\Sigma\Rightarrow\Box^{k-1}\bot$ has a derivation of height at most $h$, contradicting the inductive hypothesis with $\Lambda=\Pi,\Box\Sigma,\Sigma$ and $\ell=k-1$.
\end{proof}

\begin{lemma}
\label{lem:5-2}

Let \(j\geq0\) be an integer, and let $\Delta$ be a finite multiset whose members have the form \(\Box^iq\) with \(i>j\), where $q$ is the propositional variable we have fixed in Definition \ref{def:approximations}. Then
\[
\mathsf G\nvdash\Delta\Rightarrow\Box^jq.
\]

\end{lemma}

\begin{proof}
We prove this claim by induction on $j$.

\smallskip\noindent
\emph{Base case: $j=0$.}
Suppose for contradiction that $\Delta\Rightarrow q$ has a derivation. It cannot be initial, since $q\notin\Delta$. No propositional rule or $R_\Box$ can be its last rule-application. The only remaining possibility is $U$, whose premise $\Delta\Rightarrow\Box\bot$ is excluded by Lemma \ref{lem:5-1}. Thus, for $j=0$, $\Delta\Rightarrow q$ has no derivation in $\mathsf G$.

\smallskip\noindent
\emph{Inductive step.}
Assume the statement holds for some $j\geq0$. Suppose for a contradiction that for some finite multiset $\Delta$ whose members have the form $\Box^iq$ with $i>j+1$, the sequent $\Delta\Rightarrow\Box^{j+1}q$ has a $\mathsf G$ derivation. The inductive hypothesis states that, for every finite multiset $\Lambda$ whose members have the form $\Box^iq$ with $i>j$, the sequent $\Lambda\Rightarrow\Box^jq$ has no $\mathsf G$ derivation.

The sequent cannot be initial, and no propositional rule can be its last rule-application. The derivation cannot end in $U$, since its premise $\Delta\Rightarrow\Box\bot$ would contradict Lemma \ref{lem:5-1}. Thus the last rule must be $R_\Box$. Since every member of $\Delta$ has $\Box$ as its principal connective, write $\Delta=\Box\Sigma$. The last step is
\[
\vlderivation{
  \vlin{}{R_\Box}
    {\Box\Sigma\Rightarrow\Box^{j+1}q}
    {\vlhy{\Box\Sigma,\Sigma\Rightarrow\Box^jq}}
}
\]
Each member of $\Box\Sigma$ has the form $\Box^iq$ with $i>j+1$, and the corresponding member of $\Sigma$ is $\Box^{i-1}q$, where $i-1>j$. Thus every member of $\Box\Sigma,\Sigma$ has the form $\Box^\ell q$ with $\ell>j$. The inductive hypothesis therefore excludes the derivability of the premise $\Box\Sigma,\Sigma\Rightarrow\Box^jq$, a contradiction.
\end{proof}

\begin{corollary}\label{cor:strict-box-iteration}
For all integers $m>n\geq0$,
\[
\nvdash_{\mathrm{IEL}}\Box^mq\to\Box^nq.
\]
\end{corollary}

\begin{proof}
Taking $\Delta=\{\Box^mq\}$ and $j=n$ in Lemma \ref{lem:5-2}, we obtain $\mathsf G\nvdash\Box^mq\Rightarrow\Box^nq$. By the Hilbert--sequent correspondence in Theorem \ref{thm:2-1} and the deduction theorem, we get $\nvdash_{\mathrm{IEL}}\Box^mq\to\Box^nq$.
\end{proof}

The next two lemmas show that, in every logic $L$ with $\mathrm{IEL}^{-}\subseteq L\subseteq\mathrm{IEL}$, the formulas $\Box^nq\to r$ become strictly stronger, while the formulas $(\Box^nq\to r)\to r$ become strictly weaker, as $n$ increases.

\begin{lemma}
\label{lem:6-1}

Let $m,n$ be integers with $m>n\geq0$. If \(\mathrm{IEL}^{-}\subseteq L\subseteq\mathrm{IEL}\), then
\[
\vdash_L(\Box^m q\to r)\to(\Box^n q\to r),
\qquad
\nvdash_L(\Box^n q\to r)\to(\Box^m q\to r).
\]

\end{lemma}

\begin{proof}

The provable implication is already an IEL\(^{-}\) theorem we have proved in Lemma \ref{lem:box-iteration}, and hence belongs to \(L\).

If the reverse implication were provable in \(L\), it would be provable in IEL. In that IEL theorem, substitute \(r:=\Box^n q\). Since \(\Box^n q\) contains only \(q\), we get
\[
\vdash_{\mathrm{IEL}}
(\Box^n q\to\Box^n q)\to(\Box^m q\to\Box^n q).
\]

Since \(\vdash_{\mathrm{IEL}}\Box^n q\to\Box^n q\), modus ponens gives \(\vdash_{\mathrm{IEL}}\Box^m q\to\Box^n q\), contradicting Corollary \ref{cor:strict-box-iteration}.

\end{proof}

\begin{lemma}
\label{lem:6-2}
Let $m,n$ be integers with $m>n\geq0$. If \(\mathrm{IEL}^{-}\subseteq L\subseteq\mathrm{IEL}\), then
\[
\vdash_L
\bigl((\Box^n q\to r)\to r\bigr)
\to\bigl((\Box^m q\to r)\to r\bigr),
\]
\[
\nvdash_L
\bigl((\Box^m q\to r)\to r\bigr)
\to\bigl((\Box^n q\to r)\to r\bigr).
\]

\end{lemma}

\begin{proof}

Assume \((\Box^n q\to r)\to r\) and \(\Box^m q\to r\). By Lemma \ref{lem:6-1}, the second assumption gives \(\Box^n q\to r\), and then the first assumption gives \(r\). Applying the deduction theorem twice, we get $\vdash_{\mathrm{IEL}^-} \bigl((\Box^n q\to r)\to r\bigr) \to\bigl((\Box^m q\to r)\to r\bigr)$, so the formula is provable in every such logic $L$.

Suppose the reverse implication were provable in \(L\) and hence in IEL. Substituting \(r:=\Box^n q\) in this IEL theorem, we obtain
\[\vdash_{\mathrm{IEL}}
\bigl((\Box^m q\to \Box^n q)\to \Box^n q\bigr)
\to\bigl((\Box^n q\to \Box^n q)\to \Box^n q\bigr).\]

On the other hand, IPC proves
\[
\Box^m q\to
\bigl((\Box^m q\to\Box^n q)\to\Box^n q\bigr)
\]
and
\[
\bigl((\Box^n q\to\Box^n q)\to\Box^n q\bigr)
\to\Box^n q.
\]
By transitivity of implication, we obtain
\[
\vdash_{\mathrm{IEL}}\Box^m q\to\Box^n q,
\]
contradicting Corollary \ref{cor:strict-box-iteration}.
\end{proof}

\section{Failure of Uniform Interpolation}

We now combine the finite approximation and strictness results to prove that both uniform post-interpolation and uniform pre-interpolation fail in every logic $L$ with $\mathrm{IEL}^{-}\subseteq L\subseteq\mathrm{IEL}$.

\begin{theorem}
\label{thm:7-1}

Let \(L\) be a logic in the language $\mathcal{L}_{\Box}$ such that
\[
\mathrm{IEL}^{-}\subseteq L\subseteq\mathrm{IEL}.
\]

The formula
\[
((q\to p)\land(\Box p\to p))\land(p\to r)
\]
has no uniform post-interpolant with respect to \(p\) in \(L\), and the formula
\[
\bigl(((q\to p)\land(\Box p\to p))\land(p\to r)\bigr)\to r
\]
has no uniform pre-interpolant with respect to \(p\) in \(L\).

\end{theorem}

\begin{proof}

First consider post-interpolants. Suppose \(\chi\) is a uniform post-interpolant of the first formula with respect to \(p\). Condition (1) of Definition \ref{def:1-1} gives $\operatorname{Voc}(\chi)\subseteq\{q,r\},$ so \(\chi\) is \(p\)-free. By condition (2) of Definition \ref{def:1-1},
\[
\vdash_L
\Bigl(((q\to p)\land(\Box p\to p))\land(p\to r)\Bigr)\to\chi.
\]

Since \(L\subseteq\mathrm{IEL}\), this formula is also provable in IEL. Theorem \ref{thm:4-2} gives an \(n\geq0\) such that
\[
\vdash_{\mathrm{IEL}}(\Box^nq\to r)\to\chi.
\tag{5.1}\label{eq:5-1}
\]

On the other hand, Lemma \ref{lem:4-1} gives
\[
\vdash_{\mathrm{IEL}^{-}}
\Bigl(((q\to p)\land(\Box p\to p))\land(p\to r)\Bigr)
\to(\Box^{n+1}q\to r).
\]

This formula is also provable in \(L\). Applying condition (3) of Definition \ref{def:1-1} with the \(p\)-free formula \(\Box^{n+1}q\to r\) yields
\[
\vdash_L\chi\to(\Box^{n+1}q\to r).
\]
This is also an IEL theorem. Composing it with \eqref{eq:5-1} gives
\[
\vdash_{\mathrm{IEL}}(\Box^nq\to r)\to(\Box^{n+1}q\to r),
\]
contradicting Lemma \ref{lem:6-1}.

Next consider pre-interpolants. Suppose \(\theta\) is a uniform pre-interpolant of the second formula with respect to \(p\). Condition (1) of Definition \ref{def:1-2} gives \(\operatorname{Voc}(\theta)\subseteq\{q,r\}\), so \(\theta\) is \(p\)-free. Condition (2) of Definition \ref{def:1-2} and \(L\subseteq\mathrm{IEL}\) give
\[
\vdash_{\mathrm{IEL}}
\theta\to\Bigl(\bigl(((q\to p)\land(\Box p\to p))\land(p\to r)\bigr)\to r\Bigr).
\]

By Corollary \ref{cor:4-3}, there is an \(n\geq0\) such that
\[
\vdash_{\mathrm{IEL}}\theta\to\bigl((\Box^nq\to r)\to r\bigr).
\tag{5.2}\label{eq:5-2}
\]

Recall that the following formula is already provable in IEL\(^{-}\) by Lemma \ref{lem:4-1}, and hence is provable in \(L\):
\[
\begin{aligned}
\vdash_L\;&\bigl((\Box^{n+1}q\to r)\to r\bigr) \to\Bigl(\bigl(((q\to p)\land(\Box p\to p))\land(p\to r)\bigr)\to r\Bigr).
\end{aligned}
\]

Apply condition (3) of Definition \ref{def:1-2} with the \(p\)-free formula \((\Box^{n+1}q\to r)\to r\) to obtain
\[
\vdash_L\bigl((\Box^{n+1}q\to r)\to r\bigr)\to\theta.
\]
This is also an IEL theorem. Composing it with \eqref{eq:5-2} yields
\[
\vdash_{\mathrm{IEL}}
\bigl((\Box^{n+1}q\to r)\to r\bigr)
\to\bigl((\Box^nq\to r)\to r\bigr),
\]
contradicting Lemma \ref{lem:6-2}.

\end{proof}

\begin{corollary}
\label{cor:7-2}

Neither IEL\(^{-}\) nor IEL has the uniform interpolation property. For each logic, the existence requirements for post-interpolants and pre-interpolants fail separately.

\end{corollary}

\begin{proof}

Take \(L=\mathrm{IEL}^{-}\) and \(L=\mathrm{IEL}\) in Theorem \ref{thm:7-1} and apply Definition \ref{def:1-3}.

\end{proof}

\section{Conclusion and Future Work}\label{sec:conclusion}

Neither $\mathrm{IEL}^{-}$ nor $\mathrm{IEL}$ has uniform interpolation. The same explicit formulas witness the failure of uniform post-interpolation and uniform pre-interpolation throughout the interval $\mathrm{IEL}^{-}\subseteq L\subseteq\mathrm{IEL}$. In particular, adding weak factivity does not remove the obstruction.

The proof combines a syntactic finite approximation theorem with strictness of the resulting modal iterations. The theorem accounts for every $p$-free consequence of the post-interpolation counterexample, while strictness ensures that no finite stage captures them all. Corollary \ref{cor:4-3} gives the corresponding description of $p$-free sufficient premises for the pre-interpolation counterexample.

The positive result for intuitionistic strong L\"ob logic $\mathrm{iSL}$~\cite{FereeEtAl2024} shows that co-reflection alone does not prevent uniform interpolation. It is therefore a natural question to ask which extensions of $\mathrm{IEL}^{-}$ alter the approximation mechanism established here sufficiently to recover uniform interpolation. For each fixed \(n\geq 1\), adding \(\Box^{n+1}\varphi\to\Box^n\varphi\) to \(\mathrm{IEL}^{-}\) makes the two approximation sequences considered here stabilize after some finite step. The case \(n=1\) gives propositional lax logic (\(\mathrm{PLL}\)), for which uniform interpolation remains an open problem; see~\cite{Iemhoff2026PLL}.

\vspace{1.5em}

\textbf{Acknowledgement.}  The author completed this work during a research visit to the Institute for Logic, Language and Computation (ILLC) at the University of Amsterdam, with Dr.\ Marianna Girlando as host. The author thanks ILLC for providing an excellent working environment and is especially grateful to Dr.\ Girlando for her helpful comments, her help in checking the ideas presented in this paper, and her assistance with improving its language. The author is supported by Tsinghua University's Initiative for Advancing First-Class and World-Leading Disciplines in the Humanities and Social Sciences.

\textbf{Usage of AI.} During the preparation of this work, the author used GPT-6 Astra solely to improve the readability and language of Section 1. After using this tool, the author reviewed and edited the content as needed and takes full responsibility for the content of the paper.


\begin{thebibliography}{99}

\bibitem{ArtemovProtopopescu2016}
S. Artemov and T. Protopopescu. Intuitionistic epistemic logic. \emph{The Review of Symbolic Logic}, 9(2):266--298, 2016. \href{https://doi.org/10.1017/S1755020315000374}{doi:10.1017/S1755020315000374}.

\bibitem{Craig1957}
W. Craig. Three uses of the Herbrand--Gentzen theorem in relating model theory and proof theory. \emph{The Journal of Symbolic Logic}, 22(3):269--285, 1957. \href{https://doi.org/10.2307/2963594}{doi:10.2307/2963594}.

\bibitem{DAgostino2007}
G. D'Agostino. Uniform interpolation, bisimulation quantifiers, and fixed points. In \emph{Logic, Language, and Computation (TbiLLC 2005)}, Lecture Notes in Computer Science 4363, pages 96--116. Springer, 2007. \href{https://doi.org/10.1007/978-3-540-75144-1_8}{doi:10.1007/978-3-540-75144-1\_8}.

\bibitem{FereeEtAl2024}
H. F\'{e}r\'{e}e, I. van der Giessen, S. van Gool, and I. Shillito. Mechanised uniform interpolation for modal logics K, GL, and iSL. In \emph{Automated Reasoning (IJCAR 2024)}, Part II, Lecture Notes in Computer Science 14740, pages 43--60. Springer, 2024. \href{https://doi.org/10.1007/978-3-031-63501-4_3}{doi:10.1007/978-3-031-63501-4\_3}.

\bibitem{GhilardiZawadowski1995}
S. Ghilardi and M. Zawadowski. Undefinability of propositional quantifiers in the modal system S4. \emph{Studia Logica}, 55(2):259--271, 1995. \href{https://doi.org/10.1007/BF01061237}{doi:10.1007/BF01061237}.

\bibitem{Giessen2022}
I. van der Giessen. \emph{Uniform Interpolation and Admissible Rules: Proof-theoretic Investigations into (Intuitionistic) Modal Logics}. PhD thesis, Utrecht University, 2022. \href{https://doi.org/10.33540/1486}{doi:10.33540/1486}.

\bibitem{GiessenShillito2026}
I. van der Giessen and I. Shillito. Uniform interpolation with constructive diamond. In \emph{Automated Reasoning (IJCAR 2026)}, Part I, Lecture Notes in Computer Science 16688, pages 357--376. Springer, 2026. \href{https://doi.org/10.1007/978-3-032-32589-1_22}{doi:10.1007/978-3-032-32589-1\_22}.

\bibitem{Iemhoff2019}
R. Iemhoff. Uniform interpolation and the existence of sequent calculi. \emph{Annals of Pure and Applied Logic}, 170(11):102711, 2019. \href{https://doi.org/10.1016/j.apal.2019.05.008}{doi:10.1016/j.apal.2019.05.008}.

\bibitem{Iemhoff2026PLL}
R. Iemhoff. Proof theory for lax logic. arXiv:2209.08976, 2022; revised version, 2026. \href{https://arxiv.org/abs/2209.08976v2}{arXiv:2209.08976v2}.

\bibitem{KowalskiMetcalfe2018}
T. Kowalski and G. Metcalfe. Coherence in modal logic. In \emph{Advances in Modal Logic}, volume 12, pages 459--472. College Publications, 2018. \href{https://doi.org/10.7892/boris.119774}{doi:10.7892/boris.119774}.

\bibitem{KowalskiMetcalfe2019}
T. Kowalski and G. Metcalfe. Uniform interpolation and coherence. \emph{Annals of Pure and Applied Logic}, 170(7):825--841, 2019. \href{https://doi.org/10.1016/j.apal.2019.02.004}{doi:10.1016/j.apal.2019.02.004}.

\bibitem{KrupskiYatmanov2016}
V. N. Krupski and A. Yatmanov. Sequent calculus for intuitionistic epistemic logic IEL. In S. Artemov and A. Nerode, editors, \emph{Logical Foundations of Computer Science (LFCS 2016)}, volume 9537 of \emph{Lecture Notes in Computer Science}, pages 187--201. Springer, Cham, 2016. \href{https://doi.org/10.1007/978-3-319-27683-0_14}{doi:10.1007/978-3-319-27683-0\_14}.

\bibitem{Pitts1992}
A. M. Pitts. On an interpretation of second order quantification in first order intuitionistic propositional logic. \emph{The Journal of Symbolic Logic}, 57(1):33--52, 1992. \href{https://doi.org/10.2307/2275175}{doi:10.2307/2275175}.

\bibitem{SuSano2019}
Y. Su and K. Sano. First-order intuitionistic epistemic logic. In \emph{Logic, Rationality, and Interaction (LORI 2019)}, Lecture Notes in Computer Science 11813, pages 326--339. Springer, 2019. \href{https://doi.org/10.1007/978-3-662-60292-8_24}{doi:10.1007/978-3-662-60292-8\_24}.

\end{thebibliography}
\end{document}